\documentclass[seceq]{ptptex}

\usepackage{graphicx}

\newcommand{\tr}{\mathop{\rm tr}\nolimits}

\title{
Ginsparg-Wilson Relation and Admissibility Condition 
in Noncommutative Geometry\footnote{Talk given at Nishinomiya-Yukawa 
Memorial Symposium on Theoretical Physics 
``Noncommutative Geometry and Quantum Spacetime in Physics", Japan, Nov.11-15, 2006.
This talk is based on the work with H.Aoki and S.Iso besides my own work.}
}

\author{
Keiichi \textsc{Nagao}\footnote{email: nagao@mx.ibaraki.ac.jp }
}

\inst{
Theoretical Physics Laboratory, College of Education, Ibaraki University, 
Mito 310-8512, Japan}

\abst{

Ginsparg-Wilson relation and admissibility condition 
have the key role to construct lattice chiral gauge theories. 
They are also useful to define the chiral structure in finite 
noncommutative geometries or matrix models. 
We discuss their usefulness briefly. 
}

\begin{document}

\maketitle

\section{Introduction}

Noncommutative (NC) geometry\cite{Connes} had attracted much attention recently 
from various motivations.
Topologically nontrivial configurations in finite NC 
geometries or matrix models\cite{BFSS}\cite{IKKT} 
have been constructed based on algebraic 
K-theory and projective modules in many papers, but 
it would be better if we could obtain an index operator 
which takes an integer even at a finite cutoff, since 
we need to perform the Kaluza-Klein compactification of extra dimensions with 
nontrivial indices to construct four dimensional chiral gauge theories.
This can be realized if we utilize 
the Ginsparg-Wilson (GW) relation\cite{GinspargWilson} and 
the admissibility condition\cite{Luscher:1981zq}
\cite{Hernandez:1998et}\cite{Neuberger:1999pz}, which 
were developed in lattice gauge theory (LGT) 
to construct chiral gauge theories\cite{Luscher:1998du}.



\section{{\bf GW formulation in finite NC geometry}}
In ref.\cite{AIN2}, 
we proposed a general prescription to construct chirality 
and Dirac operators satisfying the GW relation and an index 
in general gauge field backgrounds on general finite NC geometries. 
The prescription proposed in ref.\cite{AIN2} is as follows. 
Let us introduce two hermitian chirality operators: 
one is a chirality operator $\gamma$, which is assumed to be independent of 
gauge fields, while the other is constructed in terms of a hermitian operator $H$ as 
$\hat{\gamma} \equiv \frac{H}{\sqrt{H^2}}, \ H^{\dagger}=H$.
$\gamma$ and $\hat\gamma$ satisfy $\gamma^2=\hat{\gamma}^2=1$.  
$\hat\gamma$ depends on gauge fields through $H$. 
The Dirac operator $D_{GW}$ is defined by 
$1- \gamma \hat{\gamma} = f(a,\gamma) D_{GW}$, 
where $a$ is a small parameter. 
$H$ and the function $f$ must be properly chosen 
so that the $D_{GW}$ is free of species doubling and 
behaves correctly in the commutative limit ($a \rightarrow 0$). 
$D_{GW}$ satisfies the GW relation\cite{GinspargWilson}: 
$\gamma D_{GW}+D_{GW} \hat{\gamma}=0$. 
Therefore the fermionic action 
$S_F={\rm tr}(\bar\Psi D_{GW} \Psi) \label{fermionic_action}$
is invariant under the modified chiral transformation~\cite{Luscher, Nieder,AIN2}
$\delta \Psi =i \lambda \hat{\gamma} \Psi, \, 
\delta \bar{\Psi} = i \bar{\Psi}\lambda \gamma$. 
The Jacobian, however, is not invariant and has the form 
$q(\lambda)=
\frac{1}{2}{\cal T}r(\lambda \hat{\gamma} +\lambda \gamma)$, 
where ${\cal T}r$ is a trace of operators acting on matrices. 
This $q(\lambda)$ is expected to provide a topological charge density, and 
the index for $\lambda=1$.

An index theorem is given by 
${\rm{index}}D_{GW}\equiv (n_+ - n_-)=\frac{1}{2} {\cal T}r(\gamma+\hat{\gamma})$, 
where $n_\pm$ are numbers of zero eigenstates of $D_{GW}$ with a positive 
(or negative) chirality (for either $\gamma$ or $\hat\gamma$). 
This index theorem can be easily proven\cite{AIN3}, as done in 
LGT\cite{Hasenfratzindex}\cite{Luscher}. 
The index is invariant 
under small deformation of any parameters such as 
gauge configurations in the operator $H$. 
We note that $\hat\gamma$ becomes singular when $H$ has zero modes. 
When an eigenvalue of $H$ crosses zero, the value of 
${\cal T}r \hat{\gamma}$ changes by two.

In LGT the configuration space of gauge fields is topologically
trivial if we do not impose an admissibility 
condition\cite{Luscher:1981zq,Hernandez:1998et,Neuberger:1999pz} on gauge fields. 
This condition suppresses the fluctuation of gauge fields, and 
consequently forms a topological structure composed of isolated islands 
in the configuration space. 
This condition also excludes zero modes of $H$. 
In ref.\cite{AIN2} we have thus expected that a similar mechanism would work 
also in finite NC geometries or matrix models, 
and that the index could take various integers 
according to gauge configurations. 

\section{\bf The index on fuzzy 2-sphere}

In ref.\cite{AIN2} we provided a set of simplest chirality and Dirac operators 
on fuzzy 2-sphere, as a concrete example given by the prescription.
The set in the absence of gauge fields corresponds to 
that constructed earlier in ref.\cite{balaGW}. 
The properties of $D_{GW}$ and other types of Dirac operators 
$D_{WW}$\cite{Carow-Watamura:1996wg} and $D_{GKP}$\cite{Grosse:1994ed} 
are summarized in Table I, which suggests that some kind of Nielsen-Ninomiya's 
theorem exists in matrix model or NC geometry. 
The properties of these Dirac operators are also discussed in 
ref.\cite{balagovi,balaGW}.
$D_{WW}$ has no chiral anomaly. 
The source of the chiral anomaly in $D_{GKP}$ is 
the breaking in a cut-off scale of the action under 
the chiral transformation\cite{AIN1}, and 
that in $D_{GW}$ is the Jacobian.
The nontrivial Jacobian is shown to have the correct form of 
the Chern character in the commutative limit\cite{AIN2}.


\begin{table}[htb]
\caption{The properties of three types of Dirac operators on fuzzy 2-sphere}
\begin{center}
\renewcommand{\arraystretch}{1.2}
\begin{tabular}{|c@{\quad\vrule width0.8pt\quad}l|c|c|c|}
\hline
Dirac operator &\multicolumn{2}{c|}{chiral symmetry} & no doublers & counterpart in LGT \\
\hline
$D_{WW}$ & $D_{WW} \Gamma + \Gamma D_{WW} =0$ & $\bigcirc$ & $\times$ & naive fermion \\
\hline 
$D_{GKP}$ & $D_{GKP} \Gamma + \Gamma D_{GKP} ={\cal O}(1/L)$ & $\times$ & $\bigcirc$ & Wilson fermion \\
\hline  
$D_{GW}$ & $D_{GW} \hat\Gamma + \Gamma D_{GW} =0$ & $\bigcirc$ & $\bigcirc$ & GW fermion \\
\hline 
\end{tabular}
\end{center}
\end{table}

$D_{GW}$ works well. 
The index, however, cannot take nonzero integers on fuzzy 2-sphere.
We need to apply projective modules to the index so that 
it can take nonzero integers\cite{Balachandran:2003ay}\cite{AIN3}. 
The modified index is symbolically expressed as 
${\text {index}} D_{GW} 
= \frac{1}{2}{\cal T}r \left\{ P^{(m)}[ A_\mu^{(m)}]
(\gamma + \hat{\gamma}[ A_\mu^{(m)}])\right\}=m$. 
The gauge fields $A_\mu^{(m)}$ are determined dependent on $m$.
$P^{(m)}$ is a projector to pick up a Hilbert space on which $\hat\gamma$ acts. 
The insertion of $P^{(m)}$ is necessary on fuzzy 2-sphere. 
The configuration with $m=\pm1$\cite{Balachandran:2003ay} is interpreted as 
the 't Hooft-Polyakov monopole\cite{AIN3}\cite{AIMN}\cite{Aoki:2006wv}.
As explained above, the index cannot take nonzero integers on fuzzy 2-sphere 
without the projector.
Furthermore, the naive imposition of an admissibility condition on gauge fields, 
which can be written down so that zero modes of $H$ are excluded, 
results in providing just a vacuum sector with trivial configurations. 
On a NC torus, however, the situation changes, since gauge fields on a NC torus 
are defined compactly as in LGT.

\section{\bf The index on a NC torus}

Since a NC torus\cite{CDS} has a lattice structure\cite{Ambjorn:1999ts}, 
we can use the overlap Dirac operator\cite{Neuberger:1998fp}, 
which is a practical solution to the GW relation in LGT, 
by replacing lattice difference operators with their matrix correspondences 
on the NC torus\cite{Nishimura:2001dq}.
We can also construct it by the prescription explained 
in section 2\cite{Iso:2002jc}.
The nontrivial Jacobian on the NC torus 
is shown to have the form of the Chern character 
with star-products in a weak coupling expansion\cite{Iso:2002jc} 
by utilizing a topological argument in ref.\cite{Fujiwara:2002xh}.
Parity anomaly is also calculated in ref.\cite{Nishimura:2002hw}.
On the NC torus the gauge action is given by 
$S_G=N \beta \sum_{\mu > \nu} \tr 
\left[ 1-\frac{1}{2}(P_{\mu\nu} + P_{\mu\nu}^\dag) \right] $ 
where $P_{\mu\nu}$ is the plaquette. Its explicit representation 
is given in ref.\cite{Nagao:2005st}\cite{Iso:2002jc}.
This is the TEK model\cite{EK}\cite{TEK}, which was shown to 
be a nonperturbative description of 
NC Yang-Mills theory\cite{NCMM}\cite{Ambjorn:1999ts}. 
In ref.\cite{Nagao:2005st} we formulated an admissibility condition on a NC torus.
The admissibility condition is given by 
$\Vert 1- P_{\mu\nu} \Vert < \eta_{\mu\nu} \quad {\text {for all} } \,\, \, \mu >\nu $,
where $\eta_{\mu\nu}$ are some positive parameters. 
Applying arguments in refs.\cite{Hernandez:1998et,Neuberger:1999pz} 
onto the NC torus, 
it is shown that zero modes of $H$ are excluded 
if we choose $\eta_{\mu\nu}$ properly. 
The admissibility condition implies 
$\Vert \left[ \nabla_\mu , \nabla_\nu \right] \Vert < \eta_{\mu\nu}/a^2$, 
which is the bound on the field strength. 
This becomes irrelevant in the continuum limit. 
In this sence this condition is natural.

The index can be calculated by evaluating the eigenvalues of $H$. 
Namely, the index is equal to half of the difference of 
the number of the positive eigenvalues of $H$ and that of the negative ones. 
In ref.\cite{Nagao:2005st} generating many configurations of $U_\mu$ which 
satisfies the admissibility condition, we numerically analyzed the index on 
the simplest $d=2$ dimensional NC torus, 
and found various configurations with nontrivial indices.  
Since the index is topologically 
invariant against small deformation of configurations,  
this result shows 
that a topological structure is naturally realized 
in the gauge field space by the admissibility condition, 
and that the index can take nonzero integers 
without utilizing projective modules on a NC torus.


\section{{\bf Discussions}}
GW relation and admissibility condition have an essential role in finite 
NC geometries or matrix models as well as in LGT. 
It is important to construct and investigate\cite{Aoki:2006wv} GW fermions
on various NC geometries according to the prescription\cite{AIN2}. 
It is also important to study in detail the index\cite{Aoki:2006sb} on a NC torus 
to analyze the validity of the admissibility condition proposed in ref.\cite{Nagao:2005st}. 
We hope to report some progress in these directions in the future.


\section*{Acknowledgements}
The author would like to thank the organizers of the workshop 
for their hospitality, and also the participants for fruitful discussions and 
conversations.
The work of the author is supported in part 
by Grant-in-Aid for Scientific Research 
No.18740127 from the Ministry of Education, Culture, Sports, Science and 
Technology.

%

\end{document}